\documentclass{goose-article}

\hypersetup{pdfauthor={T.W.J. de Geus}}
\header{%
  T.W.J.~de~Geus, J.E.P.~van~Duuren, R.H.J.~Peerlings, M.G.D.~Geers\\%
  Materials Science and Engineering: A, 2016, 673:551-556,  %
  \doi{10.1016/j.msea.2016.06.082}, \eprint{1603.08898}%
}

\title{%
  Fracture initiation in multi-phase materials: a statistical characterization of microstructural damage sites
}
\author[1,2]{T.W.J.~de~Geus}
\author[2]{J.E.P.~van~Duuren}
\author[2]{R.H.J.~Peerlings}
\author[2]{M.G.D.~Geers}
\affil[1]{
  Materials Innovation Institute (M2i),
  P.O.~Box~5008, 2600~GA~Delft, The~Netherlands
}
\affil[2]{
  Eindhoven University of Technology, Department of Mechanical Engineering, \protect\\ \itshape\small
  P.O.~Box~513, 5600~MB~Eindhoven, The~Netherlands
}

\contact{%
  $^*$Corresponding author: \href{mailto:t.w.j.d.geus@tue.nl}{t.w.j.d.geus@tue.nl} -- \href{mailto:tom@geus.me}{tom@geus.me}  -- \href{http://www.geus.me}{www.geus.me} %
}

\begin{document}

\maketitle

\begin{abstract}%
Understanding the microstructural influence on the failure mechanisms in multi-phase materials calls for the identification of the worst-case scenario. This necessitates a statistical approach. By performing simulations directly based on micrographs, such an approach becomes feasible. This is applied here to extract the average microstructure around damage sites.
\end{abstract}


\section{Introduction}

Multi-phase materials present an attractive blend of ductility and strength by combining two or more phases in the microstructure. The distinct mechanical properties of the phases have a strong influence on the plastic response up to failure \citep[e.g.][]{Llorca1991,Lee2004,Povirk1995}. However, the phase \textit{distribution} is one of the key drivers governing damage \citep[e.g.][]{Kang2007,Segurado2003,Williams2010}.

The role of the microstructure is not easily identified as it calls for the transparent and objective identification of the worst-case configuration. Micrographs of the material's cross-section or its fracture surface \citep{Tasan2010,Cox1974,Anderson2014} are often restricted to snapshots prior to failure and require extensive expert interpretation. On the other hand, simulations are often limited to simplified microstructures with isolated observations \citep{Williams2012,Sun2009}, because their computational costs prohibit a statistically representative analysis. Idealized models \citep{Kumar2006,DeGeus2015a} do allow a statistical treatment, but the approximation due to the idealization cannot easily be quantified.

This paper overcomes some of these obstacles by performing a statistical analysis using simulations that are directly based on a series of micrographs. The goal of this paper is two-fold. First, it is demonstrated that by employing an advanced FFT-based solver, computations on true microstructures become sufficiently efficient to render the proposed statistical approach feasible. Second, the average microstructure around damage initiation sites is characterized qualitatively and quantitatively.

We consider a two-phase microstructure, represented by an ensemble of $100$ thresholded micrographs of a dual-phase steel microstructure in which martensite and ferrite are identified, considered simply as hard and soft as described in Section~\ref{sec:model} together with the mechanical model. Sections~\ref{sec:macro} and \ref{sec:micro} present the computed macroscopic and microscopic responses. Section~\ref{sec:hotspot} performs the statistical analysis by quantification of the average probability of hard phase around damage, followed with a discussion on how this insight can be used to screen microstructures for regions that may be suspected to develop damage, in Section~\ref{sec:discussion}. This paper ends with concluding remarks and an outlook in Section~\ref{sec:conclusion}.

\section{Microstructural model}
\label{sec:model}

\begin{figure}[htp]
  \centering
  \includegraphics[width=1.\textwidth]{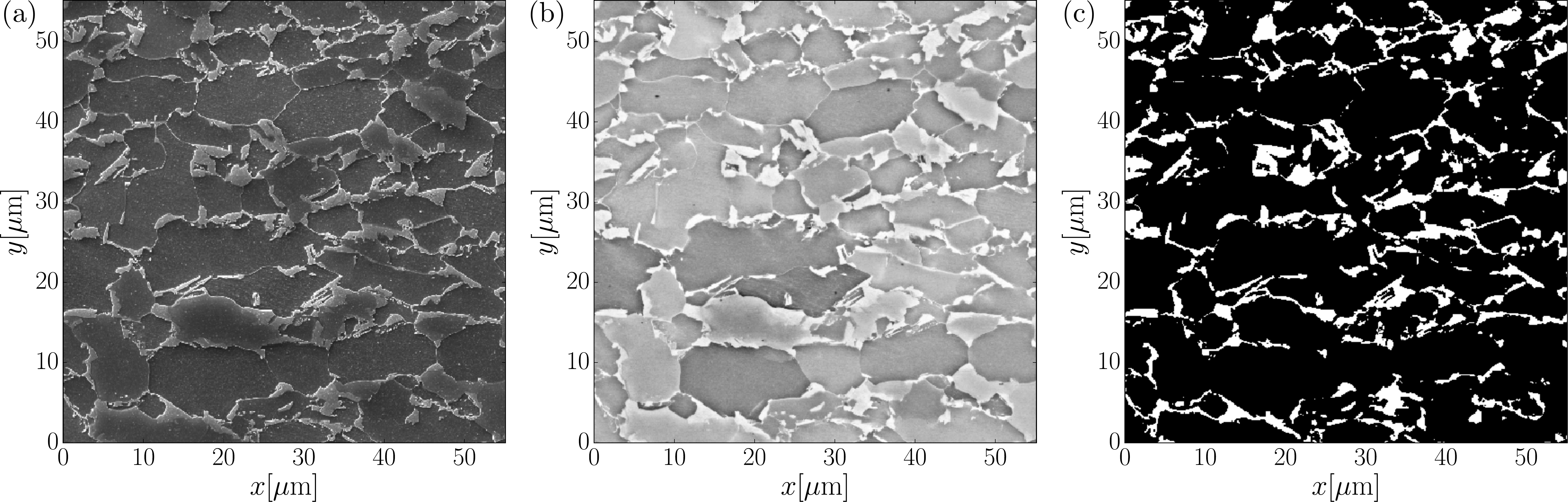}
  \caption{Typical micrographs acquired with the (a) SE detector and (b) BSE detector; (c) segmented weighted average. In each image the martensite is brighter than the ferrite. The resolution is $451 \times 451$ pixels.}
  \label{fig:microstructure}
\end{figure}

The microstructure is modeled using an ensemble of $100$ two-dimensional volume elements, each acquired from a micrograph of a commercial dual-phase steel. A protocol of grinding, polishing, and etching creates a small height difference between the hard and the soft phase. Two detectors of the scanning electron microscope are used simultaneously to image the spatial distribution of the phases. The secondary electron (SE) detector provides contrast due to the height difference (e.g.\ Figure~\ref{fig:microstructure}(a)) and the back-scatter electron (BSE) detector due to the different crystal lattice of the phases (e.g.\ Figure~\ref{fig:microstructure}(b)). The phases are identified from a weighted average of the two images. The weight is optimized to obtain a maximally separable image according to Otsu's method \citep{Otsu1979}. A typical result is shown in Figure~\ref{fig:microstructure}(c), wherein the hard phase is white and the soft phase is black.

Both phases are modeled isotropic elasto-plastic using the finite strain model as proposed by \citet{Simo1992a}. This model involves a linear relation between the logarithmic elastic strain and the Kirchhoff stress $\bm{\tau}$ which is parametrized using Young's modulus $E$ and Poisson's ratio $\nu$. The elastic domain is bounded by a yield criterion, wherein the yield stress hardens linearly with the accumulated plastic strain $\varepsilon_\mathrm{p}$ beyond an initial value $\tau_\mathrm{y0}$. The two phases are assumed elastically homogeneous and differ only in terms of their plastic response. The parameters are taken in a regime that is relevant for the dual-phase steel that has been imaged~\citep{Sun2009,Vajragupta2012}:
\begin{equation}
  \tau_\mathrm{y0}^\mathrm{hard} = 2 \tau_\mathrm{y0}^\mathrm{soft} = 0.006 \, E
  \qquad
  H^\mathrm{hard} = 2 H^\mathrm{soft} = 0.008 \, E
  \qquad
  \nu = 0.3
\end{equation}

The soft phase fails in a ductile manner, while the hard phase is assumed not to fail. A damage descriptor $D$ is used to track the material degradation, but it does not weaken the material. It is constructed such that a failure is predicted in a pixel once $D \geq 1$. Specifically, the damage descriptor is of the Johnson-Cook type \cite{Johnson1985} and incrementally compares the accumulated plastic strain rate $\dot{\varepsilon}_\mathrm{p}$ to a critical strain $\varepsilon_\mathrm{c}$ as follows
\begin{equation}
  D =
  \int_0^t \frac{\dot{\varepsilon}_\mathrm{p}}{\varepsilon_\mathrm{c} (\eta)}
  \; \mathrm{d} t^\prime
  \qquad
  \text{with}
  \qquad
  \varepsilon_\mathrm{c} = A \exp \left( - B \eta \right) + \varepsilon_\mathrm{pc}
\end{equation}
I.e.\ $\varepsilon_\mathrm{c}$ decays exponentially with the stress triaxiality $\eta$. The parameters $A$, $B$, and $\varepsilon_\mathrm{pc}$ are also taken representative for the considered steel~\cite{Vajragupta2012}:
\begin{equation}
  A = 0.2 \qquad B = 1.7 \qquad
  \varepsilon_\mathrm{pc} = 0.1
\end{equation}

The average macroscopic deformation is fully prescribed, whereas the local micro-fluctuations are the result of equilibrium and are required to be periodic (see below). Pure shear is applied according to the macroscopic logarithmic strain
\begin{equation}
\label{eq:model:pure-shear}
  \bar{\bm{\varepsilon}} =
  \tfrac{\sqrt{3}}{2} \, \bar{\varepsilon}\,
  \left(
    \vec{e}_\mathrm{x} \vec{e}_\mathrm{x} -
    \vec{e}_\mathrm{y} \vec{e}_\mathrm{y}
  \right)
\end{equation}
where $\bar{\varepsilon}$ is the macroscopic equivalent logarithmic strain. The deformation is applied in small steps of $\Delta \bar{\varepsilon} = 0.0001$, until ``macroscopic failure'' is predicted once $D \geq 1$ in $1\%$ of all pixels in the ensemble. Note that although this criterion is somewhat arbitrary, it is in the experimentally observed range, and earlier work has shown that our conclusions are insensitive to it \citep{DeGeus2015b}.

The response of each image in the ensemble is computed using an FFT-based solver \citep{Zeman2016,DeGeus2016b}. This solver is preferred over finite elements because it outperforms them in computational efficiency, both in speed and in memory footprint. The FFT relies on a regular grid, which naturally coincides with the pixels of the micrographs. Each pixel is attributed a (tensorial) degree-of-freedom which corresponds to the local deformation gradient. The material model as well as the damage are evaluated in the individual pixels. Note that because of the FFT algorithm the solution is intrinsically periodic.

\section{Macroscopic response}
\label{sec:macro}

\begin{figure}[htp]
  \centering
  \includegraphics[width=1.\textwidth]{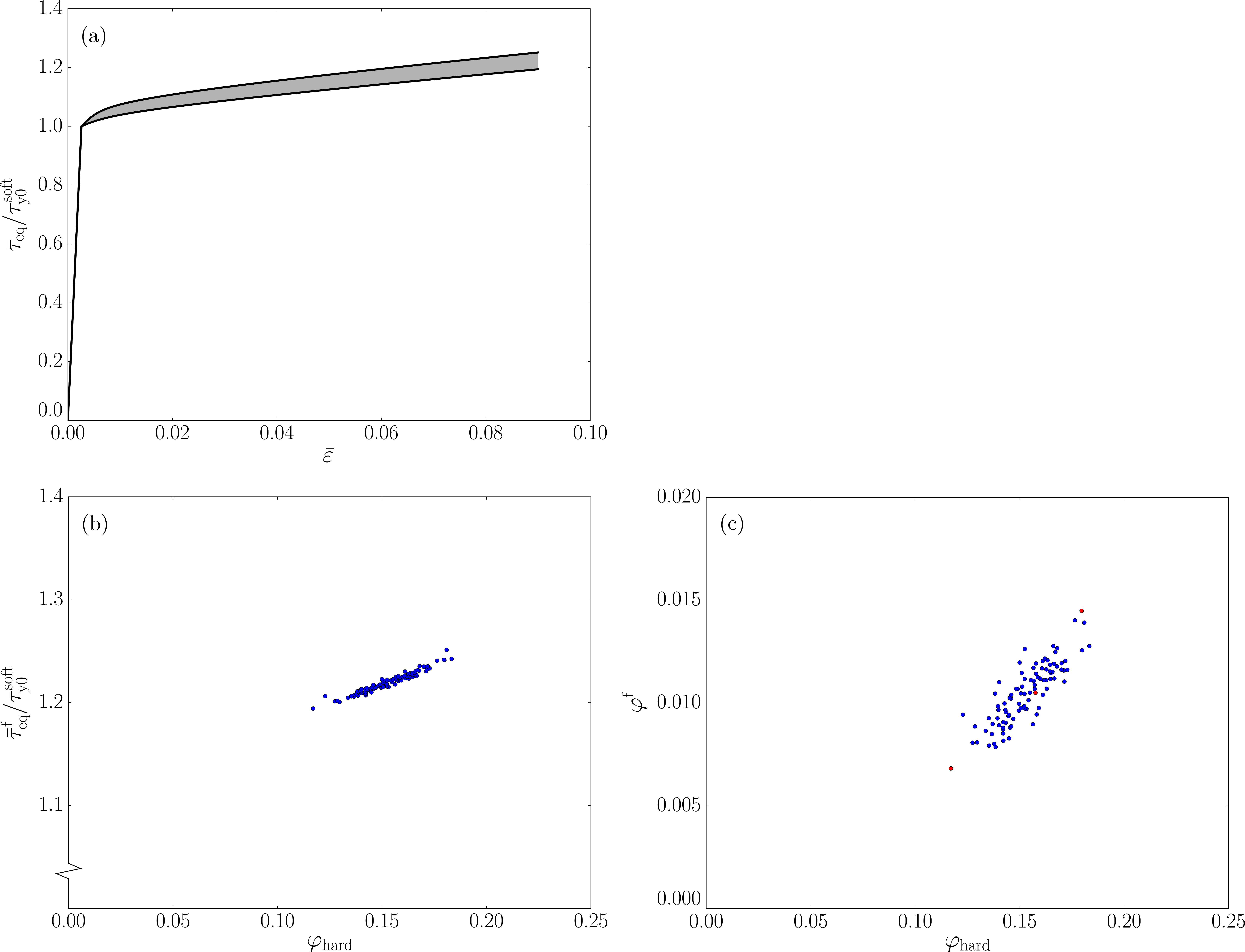}
  \caption{(a) Statistical range of macroscopic equivalent stress responses, $\bar{\tau}_\mathrm{eq}$, as a function of the applied equivalent strain $\bar{\varepsilon}$. Correlation between the hard phase volume fraction $\varphi_\mathrm{hard}$ and (b) the equivalent stress at the final increment, $\bar{\tau}_\mathrm{eq}^\mathrm{f}$ and (c) the ``void'' volume fraction $\varphi^\mathrm{f}$; for $\bar{\varepsilon} = 0.09$.}
  \label{fig:macroscopic}
\end{figure}

The range of macroscopic equivalent stress responses of the individual microstructures is shown in Figure~\ref{fig:macroscopic}(a). They initially coincide as the phases are elastically homogeneous and then start to deviate at the initial yield stress of the soft phase. The plastic response varies between different microstructures, depending on the hard phase volume fraction $\varphi_\mathrm{hard}$ of the individual microstructures. This is shown in Figure~\ref{fig:macroscopic}(b) as a strong correlation between the stress at the final increment, along the vertical axis, and $\varphi_\mathrm{hard}$ along the horizontal axis. Evidently, the precise geometrical arrangement of the phases is relatively unimportant as far as the hardening response is concerned and it is mostly the hard phase volume fraction that matters.

\section{Microscopic response}
\label{sec:micro}

\begin{figure}[htp]
  \centering
  \includegraphics[width=0.66\textwidth]{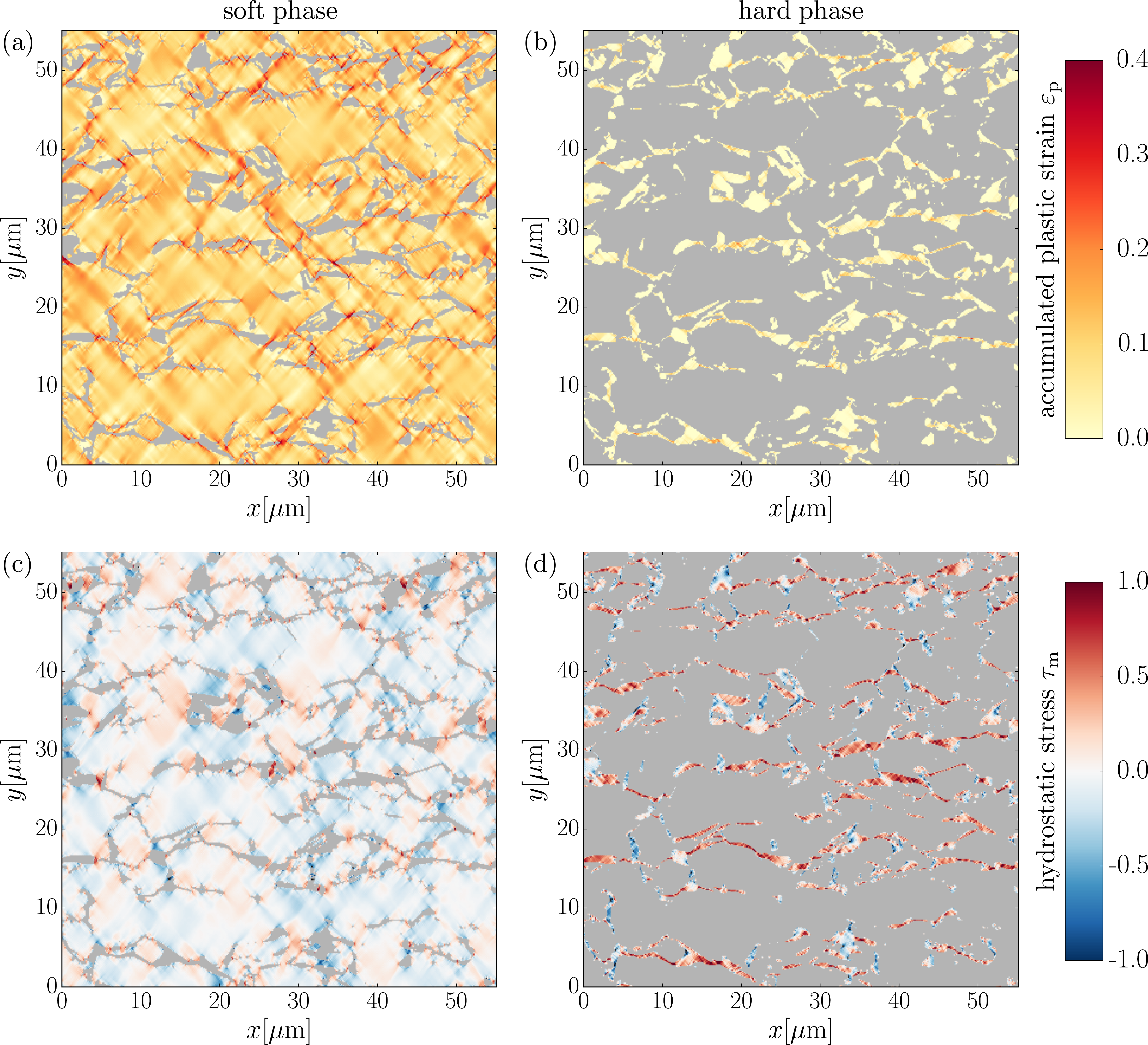}
  \caption{Typical (a--b) accumulated plastic strain and (c--d) hydrostatic stress response, for (a,c) the soft phase and (b,d) the hard phase; for $\bar{\varepsilon} = 0.09$.}
  \label{fig:typical}
\end{figure}

A typical computed microscopic response is shown in Figure~\ref{fig:typical}. The accumulated plastic strain is significantly higher in the soft phase (Figure~\ref{fig:typical}(a)) than in the hard phase (Figure~\ref{fig:typical}(b)). It is thereby localized in bands at $\pm 45$ degrees, aligned with the applied shear. The highest values are found close to the hard phase.

The hydrostatic stress is on average higher in the hard phase (Figure~\ref{fig:typical}(d)) than in the soft phase (Figure~\ref{fig:typical}(c)). In the hard phase a tensile stress (red) is observed in horizontally aligned regions of hard phase, while in the soft phase high values are observed left and right of the hard phase. The equivalent stress distribution (not shown) is qualitatively similar to the effective plastic strain response, although the equivalent stress is approximately two times higher in the hard phase than in the soft phase due to the factor of two difference in yield stress between the phases.

``Failed'' pixels are those pixels where $D \geq 1$. Their volume fraction is denoted $\varphi^\mathrm{f}$. Figure~\ref{fig:macroscopic}(c) shows $\varphi^\mathrm{f}$ as a function of the hard phase volume fraction $\varphi_\mathrm{hard}$, at the final state of all simulations. A mild correlation is observed; $\varphi^\mathrm{f}$ is typically higher for higher $\varphi_\mathrm{hard}$, but at the same time $\varphi^\mathrm{f}$ varies with up to a factor two for the same hard phase volume fraction. This implies that the damage depends much more on the precise arrangement of the phase than the hardening -- cf.\ Figure~\ref{fig:macroscopic}(b).

\begin{figure}[htp]
  \centering
  \includegraphics[width=1.\textwidth]{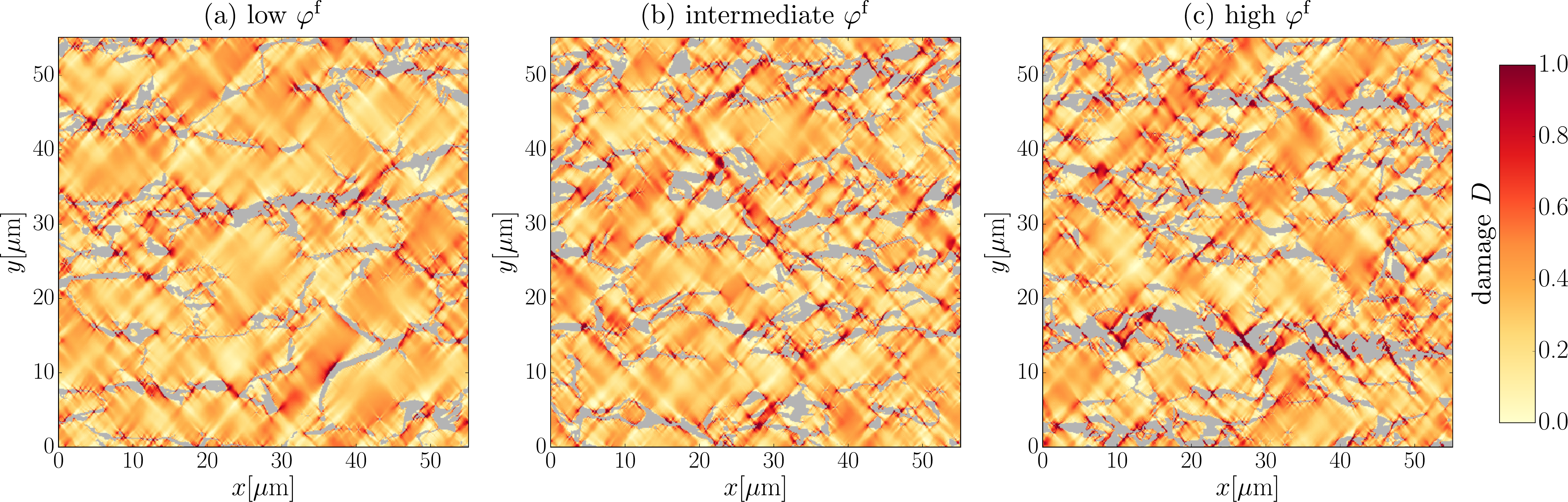}
  \caption{Typical predicted damage responses for the micrographs with the (a) lowest, (b) median, and (c) highest ``void'' volume fraction $\varphi^\mathrm{f}$; for $\bar{\varepsilon} = 0.09$.}
  \label{fig:typical:D}
\end{figure}

Three typical damage responses are shown in Figure~\ref{fig:typical:D}: the one with the lowest $\varphi^\mathrm{f}$, the median, and the one with the highest $\varphi^\mathrm{f}$. It is clear that the damage in the soft phase is determined by the adjacent presence of the hard phase. It is especially high where the soft phase is neighbored left and right by hard phase.

\section{Damage hot-spot}
\label{sec:hotspot}

Our ability to simulate \textit{many} microstructures may be exploited to quantify the role of the phase distribution on fracture initiation. For this we follow our earlier developed approach \cite{DeGeus2015a}. It is briefly explained here on the basis of a single image.

The phase distribution is described by an indicator function
\begin{equation}
  \mathcal{I} ( \vec{x}_i ) =
  \begin{cases}
    1 & \text{if}\; \vec{x}_i \in \text{hard} \\
    0 & \text{if}\; \vec{x}_i \in \text{soft} \\
  \end{cases}
\end{equation}
where the position vector $\vec{x}_i$ indicates the position of the individual pixels. The spatial average of $\mathcal{I}$ equals the hard phase volume fraction $\varphi_\mathrm{hard}$ of the microstructure. Likewise, a fracture indicator $\mathcal{D}$ is obtained which equals one where $D \geq 1$ and zero elsewhere.

The average probability of hard phase at a certain position $\Delta \vec{x}$ relative to a fractured pixel can now be computed as
\begin{equation}
  \mathcal{I}_\mathcal{D} ( \Delta \vec{x} )
  =
  \frac{
    \sum_i \mathcal{D} (\vec{x}_i)\; \mathcal{I} ( \vec{x}_i + \Delta \vec{x} )
  }{
    \sum_i \mathcal{D} (\vec{x}_i) \hfill
  }
\end{equation}
wherein $i$ loops over all pixels in the microstructure. The corresponding ensemble average $\langle \mathcal{I}_\mathcal{D} \rangle ( \Delta \vec{x} )$ is obtained by furthermore averaging over all images in the ensemble. Its value may be interpreted by comparing it to the ensemble average of the hard phase volume fraction $\langle \varphi_\mathrm{hard} \rangle$. A value $\langle \mathcal{I}_\mathcal{D} \rangle > \langle \varphi_\mathrm{hard} \rangle$ corresponds to an elevated probability of finding hard phase. Since the microstructure comprises just two phases, $\langle \mathcal{I}_\mathcal{D} \rangle < \langle \varphi_\mathrm{hard} \rangle$ corresponds to an elevated probability of finding the soft phase.

\begin{figure}[htp]
  \centering
  \includegraphics[width=.5\textwidth]{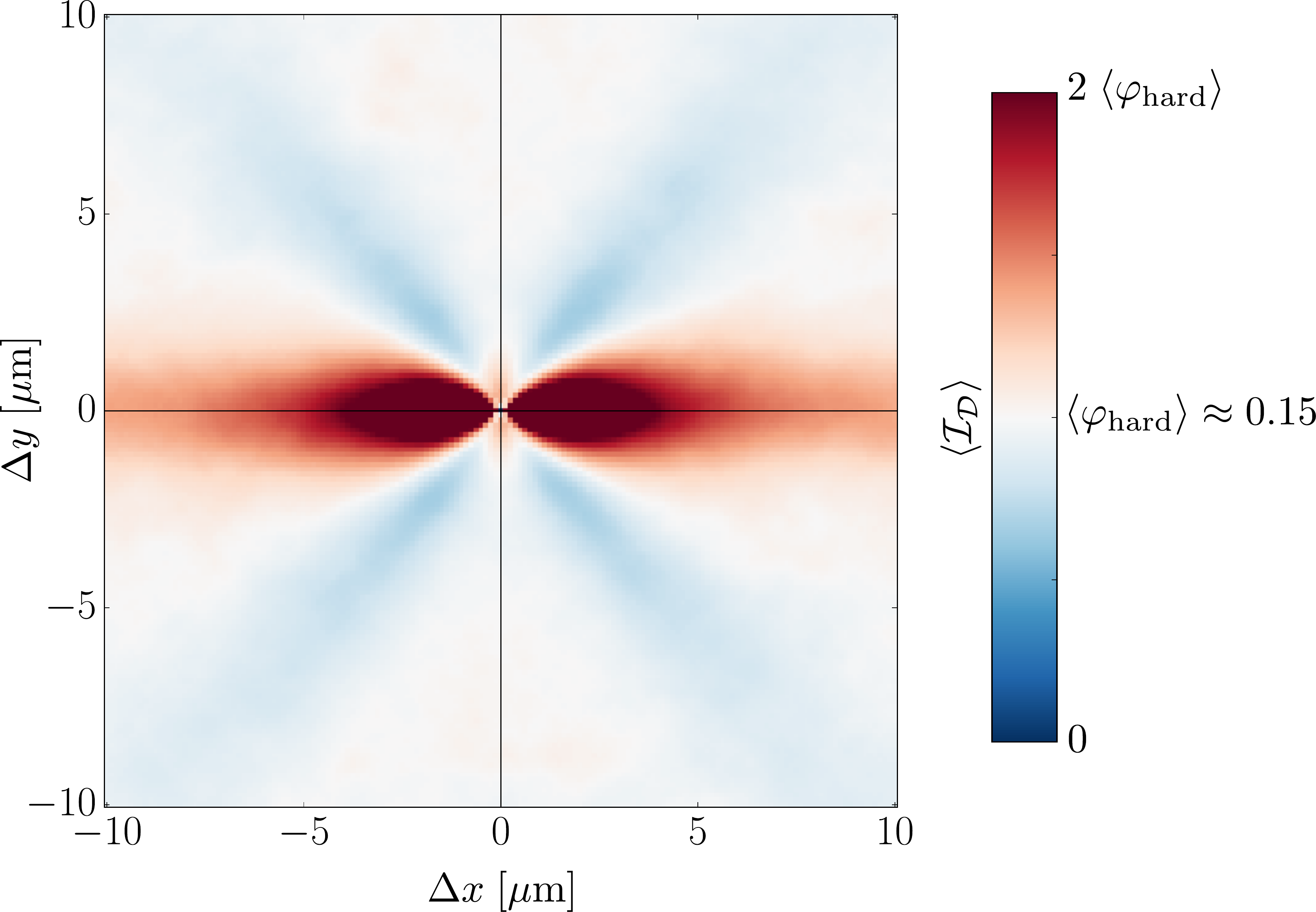}
  \caption{Probability of hard phase around damage, based on $100$ simulated microstructures. The colors indicate an elevated probability of hard phase (red) and soft phase (blue).}
  \label{fig:hotspot}
\end{figure}

The result is shown in Figure~\ref{fig:hotspot}, in which white corresponds to $\langle \varphi_\mathrm{hard} \rangle$, i.e.\ no correlation between the hard phase and fracture initiation, red corresponds to an elevated probability of the hard phase and blue to an elevated probability of the soft phase. Clearly there is quite a strong correlation with the presence of hard phase: fracture statistically initiates where a region of the hard phase aligned with tension (horizontal) is intersected by the soft phase aligned with the directions of shear (at $\pm 45$ degrees). This confirms the result by De Geus et al.\ \citep{DeGeus2015a} where a highly simplified microstructural model was used comprising equi-sized square cells in which the two phases are randomly distributed. It was argued there that the configuration of Figure~\ref{fig:hotspot} is critical because the hard phase triggers hydrostatic tension while the soft phase promotes plasticity.

A similar analysis has been performed directly on gray-scale micrographs taken after deformation (without thresholding) \citep{DeGeus2016}, on the same steel grade as used for the present study. The present results match these observations quite well. The relative position and size of the regions of the hard phase are comparable. In \citep{DeGeus2016} the regions of the soft phase at $\pm 45$ degree angles were less pronounced. Based on the present results it is observed that this correlation is less strong than the correlation with the hard phase. It is therefore likely lost in the more noise sensitive experimental setting, or due to the fact that reality is three-dimensional. This advocates the added value of performing the simulation, as these features do give insight in the mechanics of the failure mechanisms.

\section{Discussion}
\label{sec:discussion}

Having the ``footprint'' of damage-prone areas (i.e.\ Figure~\ref{fig:hotspot}) in hand, a microstructure can been screened for local phase distributions which match it. For this we compute the correlation, $\mathcal{P}_\mathcal{D} (\vec{x}_i)$ between the microstructure around the pixel at $\vec{x}_i$ pixel, characterized by $\mathcal{I} ( \vec{x}_j )$, with the probability of finding hard phase around damage, $\langle \mathcal{I}_\mathcal{D} \rangle (\Delta \vec{x})$, as follows:
\begin{equation}\label{eq:cor:predict}
  \mathcal{P}_\mathcal{D} ( \vec{x}_i ) =
  \frac{1}{N}
  \sum\limits_{j=1}^N
  \Big[
    \langle \mathcal{I}_\mathcal{D} \rangle
    ( \vec{x}_j - \vec{x}_i ) - \langle \varphi_\mathrm{hard} \rangle
  \Big]
  \;
  \Big[
    \mathcal{I} ( \vec{x}_j ) - \varphi_\mathrm{hard}
  \Big]
\end{equation}
wherein $N$ is the number of pixels.

\begin{figure}[htp]
  \centering
  \includegraphics[width=1.\textwidth]{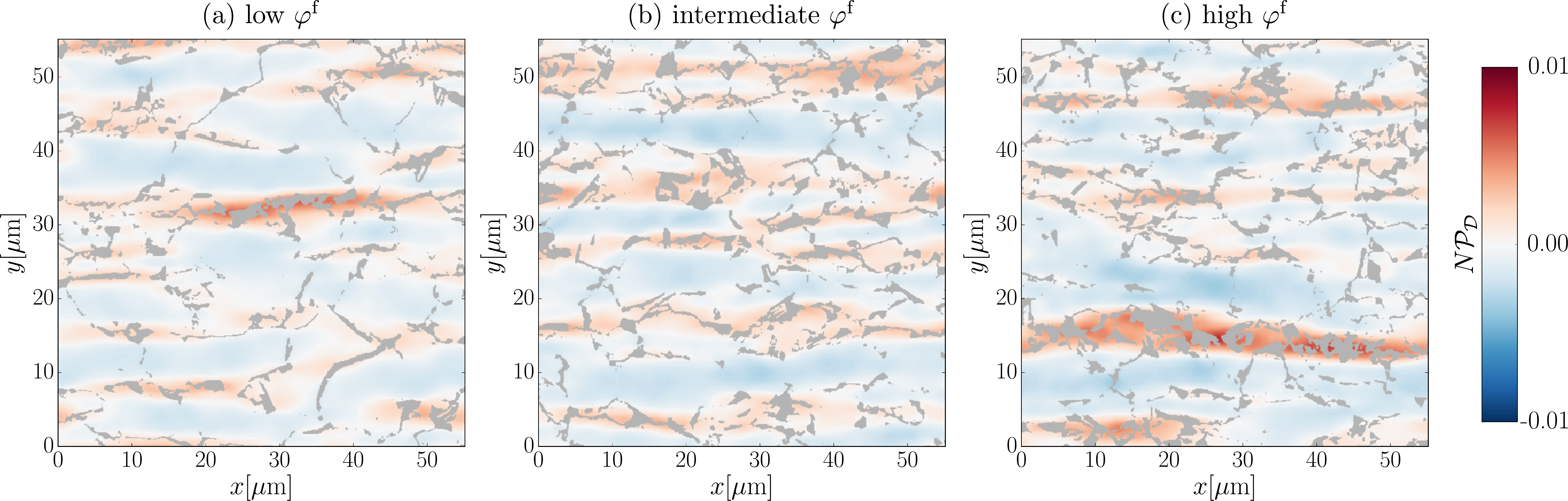}
  \caption{Damage prediction based on the probability of hard phase around fracture initiation.}
  \label{fig:correlation:hotspot}
\end{figure}

Figure~\ref{fig:correlation:hotspot} shows the result for the three microstructure of Figure~\ref{fig:typical:D}, whereby red corresponds to a substantial probability that damage will develop. The prediction is rather accurate compared with the computed damage intensities shown in Figure~\ref{fig:typical:D}, in particular in the microstructures where the damage is highest (Figures~\ref{fig:correlation:hotspot}(b,c)).

\section{Concluding remarks \& outlook}
\label{sec:conclusion}

By pairing micrographs with an FFT-based solver, a high resolution statistical analysis of fracture initiation in a dual-phase steel has been enabled. The following observations have been made:
\begin{itemize}
  \item The macroscopic hardening response scales with the hard phase volume fraction. It depends only weakly on the microstructural morphology.
  \item The microscopic response is highly dependent on the phase distribution. Plastic strain and hydrostatic stress in the soft phase localize in regions near the hard phase. High hydrostatic tensile stresses in the hard phase are observed in grains or collections of grains aligned with the direction of tension.
  \item Fracture initiation correlates weakly with the hard phase volume fraction, while a strong correlation with the local microstructural morphology has been observed.
  \item The average phase distribution around fracture initiation has been characterized. Fracture initiation is likely to occur where regions of the hard phase are aligned with the direction of tension, intersected by regions of soft phase aligned with the directions of shear. Based on these findings, it has been shown that the microstructure can be screened for damage by comparing the morphology to this critical configuration.
  \item Simulations based on real microstructures confirm observations made using a coarse idealized microstructural model (e.g.\ \citep{DeGeus2015a}).
\end{itemize}

The presented approach can be used in the future to perform high resolution quantitative comparisons of damage events in two-phase microstructures, to study the effect of morphological changes induced for example by different heat treatments. The key added value of the current analysis is that this can be done in a statistical sense and with great detail, which is desirable in the case of damage. To this end an experimental analysis has been initiated in which both the microstructural morphology as well as the damage are measured using a large set of high resolution micrographs. Paired with the proposed numerical analysis this may enable quantitative predictions.

\section*{Acknowledgments}

This research was carried out under project number M22.2.11424 in the framework of the research program of the Materials innovation institute M2i (\href{http://www.m2i.nl}{www.m2i.nl}).

\bibliography{library}

\end{document}